\documentclass[12pt]{iopart}
\usepackage{iopams}
\usepackage{graphicx}
\usepackage{iopams}
\usepackage{amssymb}
\usepackage{xcolor}
\usepackage{subfigure}
\usepackage{makecell}

\usepackage[utf8]{inputenc}
\usepackage[colorlinks=true,
            linkcolor=blue,
            citecolor=blue,
            urlcolor=blue]{hyperref}

\begin{document}
\title{Bond percolation in distorted simple cubic and body-centered cubic lattices}
\author{Bishnu Bhowmik$^1$, Sayantan Mitra$^2$, Robert M. Ziff$^3$, and Ankur Sensharma$^1$}
 \address{$^1$Department of Physics, University of Gour Banga, Malda - 732103, West Bengal, India.}
 \address {$^2$Department of Physical Sciences, Indian Institute of Science Education and Research Kolkata, Mohanpur, 741246 India.}
 \address {$^3$Center for the Study of Complex Systems and Department of Chemical Engineering, University of Michigan, Ann Arbor, Michigan 48109-2800, USA}

\date{\today}

\begin{abstract}
We investigate the effect of structural distortion on bond percolation in simple cubic and body-centered cubic lattices using extensive Monte Carlo simulations. Distortion is introduced through controlled random displacements of lattice sites, thereby modifying nearest-neighbor distances. Bond occupation is permitted only when the bond length is smaller than a prescribed connection threshold, directly coupling geometric disorder to connectivity. We find that when the connection threshold exceeds the nearest-neighbor distance of the undistorted lattice, the percolation threshold increases monotonically with distortion strength, indicating a systematic suppression of spanning. In contrast, this monotonic behavior breaks down when the connection threshold is below the nearest-neighbor distance of the undistorted lattice, highlighting a nontrivial interplay between geometric distortion and connectivity. Finite-size scaling analysis is employed to determine percolation thresholds in the thermodynamic limit and to provide evidence of universality through data collapse consistent with the standard three-dimensional critical exponent $\nu$. We further identify critical values of the connection threshold and the distortion amplitude required for global spanning when all the allowed bonds are occupied. All qualitative behaviors remain robust across both lattice geometries. These results clarify how geometric disorder reshapes percolation in three-dimensional crystalline networks.
\end{abstract}

\maketitle

\section{Introduction}
Since its introduction by Broadbent and Hammersley in 1957\cite{Broadbent}, percolation has become one of the most fundamental models in statistical physics \cite{Stauffer1994}, known for displaying nontrivial critical phenomena \cite{Albano1,Takeuchi,Grassberger}. Its simplicity makes it widely adaptable, and it has been used extensively to investigate a variety of physical processes. Notably, percolation concepts have contributed to the understanding of metal–insulator transitions \cite{Ball}, diamagnetic to ferromagnetic transitions \cite{Golovina}, unpolarized-polarized transition \cite{Aikya_2025}, etc. The influence of percolation theory extends far beyond physics. Researchers in many disciplines have adopted the model to interpret diverse natural phenomena, such as the spread of wildfires in forests \cite{Albano1}, the transport of fluids in porous media \cite{Residori}, and the transmission dynamics of infectious diseases \cite{Roman}. Numerous other complex processes have also been successfully explored through the lens of percolation \cite{Kalisky,Dotsenko,Oliveria,Abir,Soumya,Sayantan4,Vijay,Deng,Xu}.\\

Percolation theory deals with the emergence of large-scale connectivity in disordered media and comes in two fundamental forms : site percolation and bond percolation \cite{Wang}. In the site version, each lattice site can be either empty or occupied, while in the bond version the same applies to the links between neighboring sites. Beginning with an empty lattice, sites (or bonds) are randomly selected and occupied with probability $p$. As $p$ increases, occupied elements begin to form connected groups, or clusters, built from neighboring occupied sites or bonds. With further occupation the clusters grow, and at a characteristic occupation probability—known as the percolation threshold $p_c$ \cite{Derrida1985} for sites (or $p_\mathrm{b}$ for bonds)—a connected path spans the system from one side to the other. For an infinite lattice, the size of this spanning cluster diverges, signaling a continuous phase transition. Identifying this threshold and extracting the associated critical exponents \cite{Sykes1974,Sur1976} that describe the transition constitute central objectives in the study of percolation phenomena.\\

Beyond the conventional site and bond percolation, several extended variants capture richer physical scenarios. Site–bond percolation \cite{Coniglio} mixes the two mechanisms by assigning independent probabilities to both sites and bonds, allowing more flexible control of connectivity. Continuum percolation \cite{Martins} removes the lattice entirely, considering randomly placed objects such as discs, spheres, or rods that form clusters when they overlap. Directed percolation \cite{Takeuchi,Badie} imposes a preferred direction for connections, modeling processes like fluid flow, epidemic spreading, or information transfer where backflow is forbidden. Bootstrap percolation \cite{Adler,Baxter,Ming} introduces an activation rule : a site becomes stable only if a minimum number of neighbors are occupied, mimicking jamming or failure cascades. Explosive percolation \cite{Achlioptas,Riordan} modifies the growth rules to delay the formation of a spanning cluster, often producing an apparently abrupt transition. Long-range percolation \cite{Gori2017,Xun1,Xun2,Xun3} allows connections between distant sites with a probability that decays with distance, leading to small-world behavior. Correlated percolation \cite{Cao2012} incorporates spatial correlations in the disorder, relevant for real materials where occupancy is not independent. Invasion percolation \cite{Invasion1996} models fluid displacement in porous media by occupying the weakest available bond or site, generating fractal structures without tuning a probability. Continuum anisotropic percolation \cite{Singh2025} considers objects with orientation, such as needles or rods, relevant to liquid crystals or fiber networks. Together, these diverse variants extend percolation theory to complex, interacting, and realistic environments across physics, biology, and network science. For comprehensive reviews of percolation theory and its applications, see Refs. \cite{Saberi1, Mello,Hassan2}. Another important generalization is rigidity percolation\cite{Donald,Zhang}, where the percolation transition corresponds to the onset of global mechanical rigidity rather than simple connectivity. In recent times, site and bond percolation with distance-dependent connection criteria were introduced in distorted lattices \cite{Sayantan1,Sayantan2,Sayantan3,Bishnu1} in two and three dimensions.

\begin{figure}
\centering
\includegraphics[width=\linewidth, trim=0 2.5cm 0 0, clip]{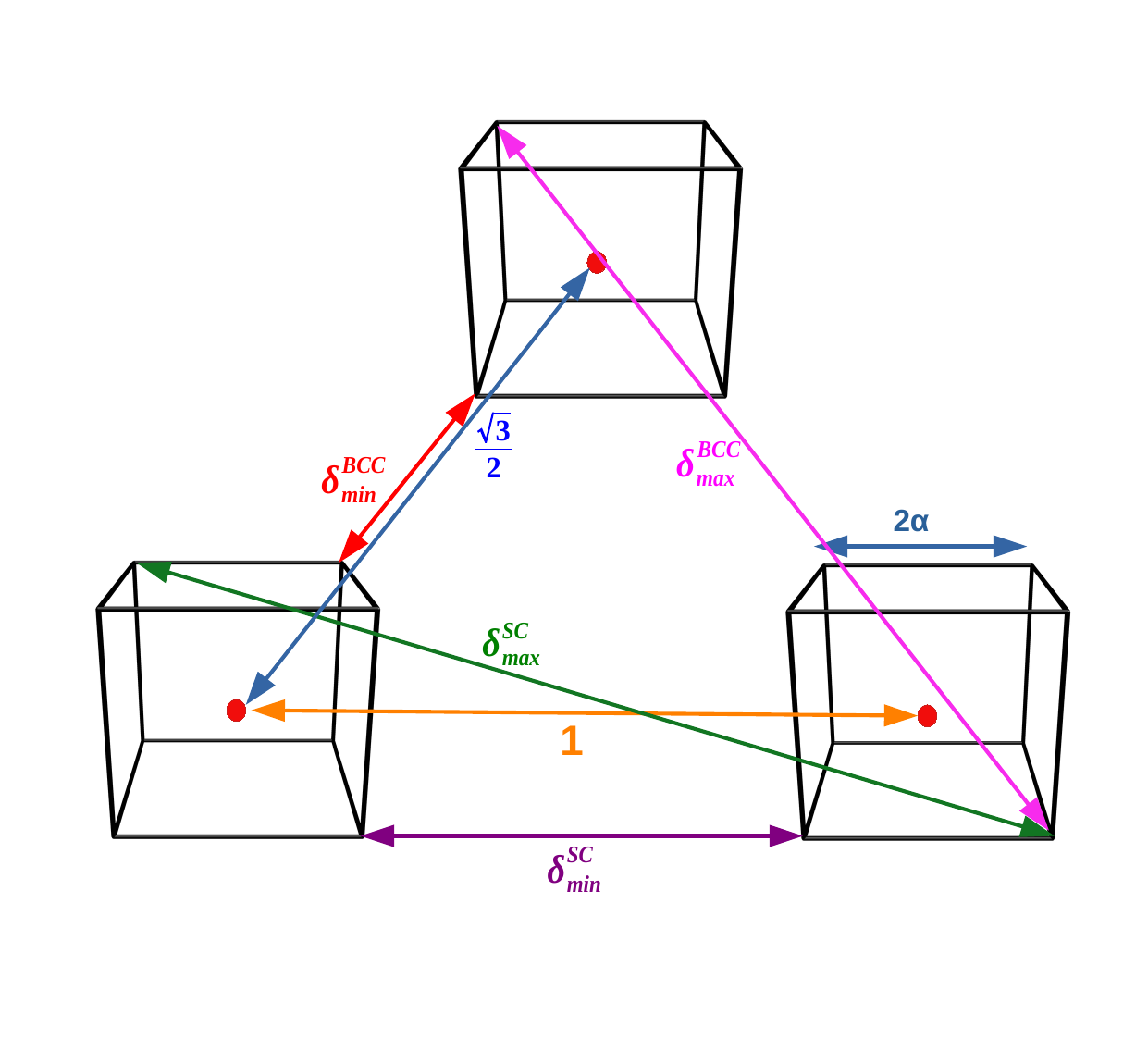}
\caption{Mechanism of distortion of simple cubic (SC) and body-centered cubic (BCC) lattices of lattice constant $1$. Red points indicate the positions of the sites in the undistorted lattice. Black cubes of length $2\alpha$ around the red lattice points indicate the possible regions of dislocated sites. The nearest neighbor distances of the undistorted lattices and the minimum and the maximum nearest-neighbor distances of corresponding distorted lattices (see Eq. \ref{dis_lattice}) are shown in the figure.}
\label{fig:sc_bcc_mech}
\end{figure}

In this paper, we adopt a distance-dependent bond occupation criterion to study bond percolation in distorted simple cubic (DSC) and distorted body-centered cubic (DBCC) lattices. Starting from perfectly ordered lattices, we introduce random positional distortions at every site, which generate a range of nearest-neighbor distances instead of a single fixed value. Consequently, each site experiences a local environment characterized by minimum and maximum bond lengths. Bonds are then randomly occupied only if they satisfy the distance-dependent occupation rule : a bond is considered eligible for occupation when its length $\delta \le d$, where 
$d$ is a prescribed connection threshold. This framework allows us to analyze how geometric distortion and connection criteria on variable bond lengths influence the percolation behavior in three-dimensional systems.

The remainder of the paper is organized as follows. In Section \ref{MM}, we introduce the model and describe the methods employed throughout the study. Section \ref{VBPT} presents the variation of the bond percolation threshold for DSC and DBCC lattices under different model parameters. In Sections \ref{CCT} and \ref{CDP}, we discuss the results derived from the analysis of the model’s critical parameters, highlighting their implications for percolation behavior in distorted lattices.

\section{The Model and the Method}\label{MM}
The mechanism of distortion is illustrated in Fig.~\ref{fig:sc_bcc_mech}. Here, $\alpha$ represents the distortion parameter. We have taken the lattice constant equal to $1$ for both simple cubic (SC) and body-centered cubic (BCC) lattices. As a result, the nearest neighbor distance for SC lattice is $1$ and for BCC lattice it is $\frac{\sqrt{3}}{2}$. Every lattice point is randomly displaced from its regular position within a cube of side $2\alpha$. After distortion, the distance between the displaced lattice points is denoted by $\delta$. The minimum $(\delta_m)$ and the maximum $(\delta_M)$ possible values of $\delta$ for distorted lattice are

\begin{eqnarray}
\delta_\mathrm{min}^\mathrm{SC} &=& 1-2\alpha \nonumber \\
\delta_\mathrm{max}^\mathrm{SC} &=& \sqrt{1+4\alpha+12\alpha^2} \nonumber \\
\delta_\mathrm{min}^\mathrm{BCC} &=& \sqrt{3}\left(\frac{1}{2}-2\alpha\right) \nonumber \\
\delta_\mathrm{max}^\mathrm{BCC} &=& \sqrt{3}\left(\frac{1}{2}+2\alpha\right)
\label{dis_lattice}
\end{eqnarray}

Since this study focuses on bond percolation, all lattice sites are assumed to be occupied. A bond is considered occupied only when its length 
$(\delta)$ is less than or equal to the connection threshold $d$. We simulate bond percolation on DSC and DBCC lattices of linear dimension $L$ to determine the bond percolation threshold $p_\mathrm{b}(\alpha,d)$ as a function of the lattice distortion parameter $\alpha$ and the connection threshold $d$. The simulation procedure for each value of $\alpha$ and $d$ is as follows :

\begin{enumerate}
\item \textbf{Lattice generation :} The first task is to generate the SC and BCC lattices of linear dimension $L$ with proper site placement to maintain lattice constant equals to $1$. In addition, it is necessary to identify all ten neighboring bonds and all fourteen neighboring bonds of each bond for SC and BCC lattice respectively.

\item \textbf{Lattice distortion :} After generating the SC and BCC lattices, each lattice site is displaced from its original coordinate, say, $(x,y,z)$ to introduce geometric distortion into the system. For every site, we draw three independent random numbers $(r_x,r_y,r_z)$, each uniformly distributed within the interval $[-\alpha,+\alpha]$. These random values represent small shifts along the $x$, $y$, and $z$ directions, respectively. The new, distorted position of the site is then obtained by adding these displacements to the original coordinates, resulting in the shifted position $(x+r_x,y+r_y,z+r_z)$. This procedure ensures that every site in the lattice is perturbed independently, producing a spatially disordered structure in which nearest-neighbor distances vary from site to site. The parameter $\alpha$ controls the strength of the distortion : larger values of $\alpha$ lead to greater deviations from the ideal lattice geometry and hence a broader distribution of neighbor distances. This controlled distortion plays a crucial role in determining which bonds satisfy the distance-dependent occupation criteria used in this percolation analysis.

\item \textbf{Bond occupation process :} In the distorted lattices, the bond length $\delta$ no longer remains fixed but varies between a minimum and a maximum value determined by the chosen distortion parameter $\alpha$, as given in Eq.~(\ref{dis_lattice}). As $\alpha$ increases, the minimum bond length decreases due to sites shifting closer together, while the maximum bond length increases as some sites move farther apart. To regulate bond occupation under this geometric disorder, we introduce a distance $d$, referred to as the connection threshold. A bond is considered eligible for occupation only if it satisfies the connection criterion $\delta \le d$. This threshold-based rule ensures that bond formation depends directly on the distorted local geometry of the lattice. In simulation, the bond occupation process is as follows :
\begin{itemize}
\item A bond occupation order is set at the beginning of the percolation process.
\item For both DSC and DBCC lattices, the bond lengths $\delta$ are computed for all bonds in the predefined occupation sequence. These calculated distances are then used to determine which bonds satisfy the connection criterion and can be considered for occupation during the percolation process.
\item If \(\delta <d\), the bond is occupied. Otherwise, the bond is rejected and another one is selected.
\end{itemize}
    \item \textbf{Cluster analysis :} After each successful bond occupation, the newly occupied bond is added to its respective cluster. A check is performed to determine if a spanning cluster has been formed. A spanning cluster is defined as a pathway of occupied bonds that connects opposite sides of the lattice.
    \item \textbf{Repeat until spanning :} Steps $4$ and $5$ are repeated until a spanning cluster is found.
    
    \item \textbf{Fraction of occupied bonds :} The fraction of occupied bonds, \(f_\mathrm{b}\), is calculated by dividing the number of occupied bonds by the total number of possible bonds.
    \item \textbf{Multiple realizations :} Steps $4$ through $7$ are repeated for $1000$ independent realizations using the same values for \(\alpha \) and \(d\).
    \item \textbf{Threshold calculation :} The average value of \(f_{b}\) over the $1000$ realizations is calculated to determine the bond percolation threshold, \(p_\mathrm{b}(\alpha ,d)\) for a lattice of linear size $L$.
\end{enumerate}
The cluster numbering and spanning analysis have been done by the Newman-Ziff algorithm \cite{Newman2}.

\begin{figure*}
	  \subfigure[]{\includegraphics[scale=0.37]{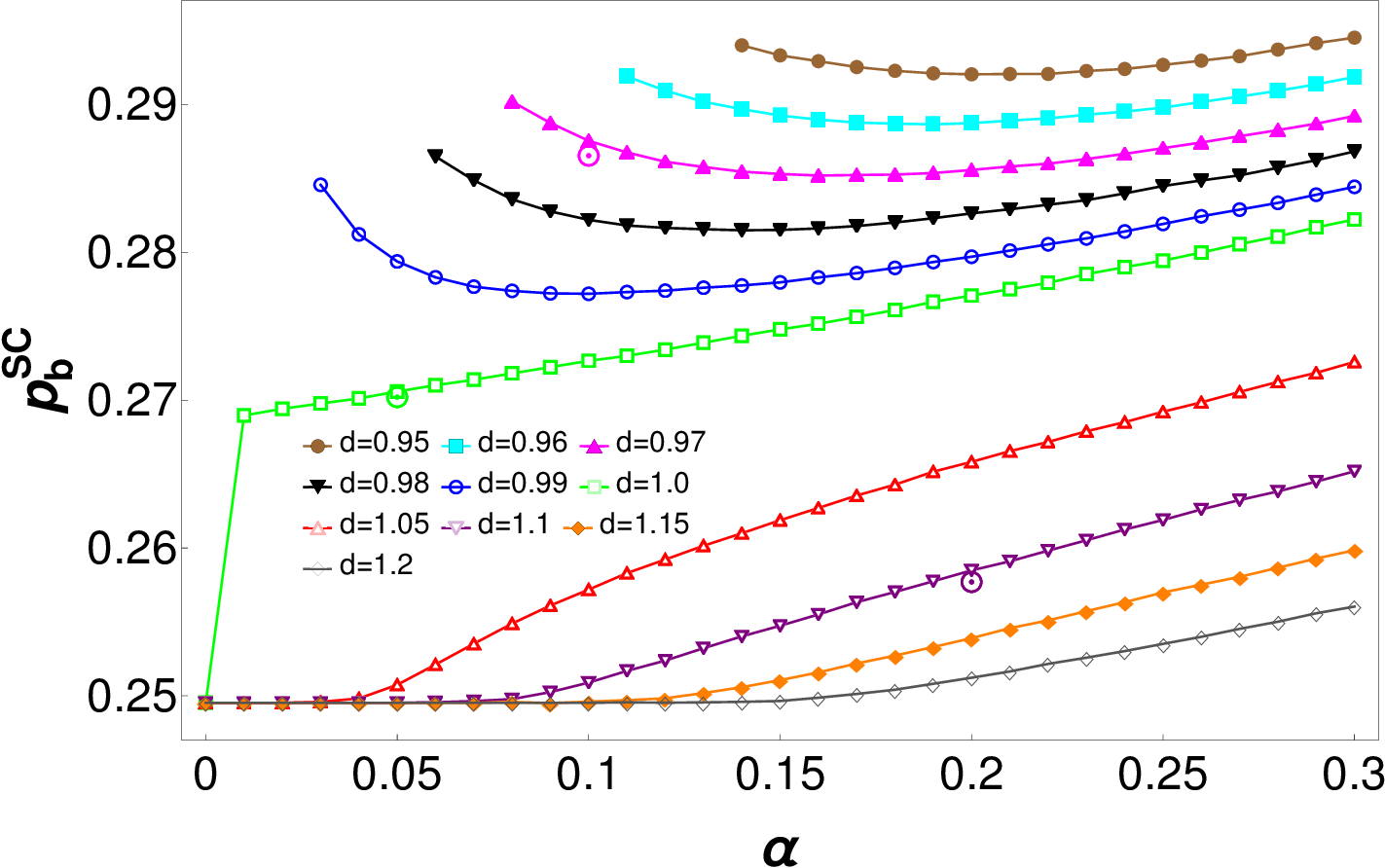}}\hspace{0.2cm}
      \subfigure[]{\includegraphics[scale=0.47]{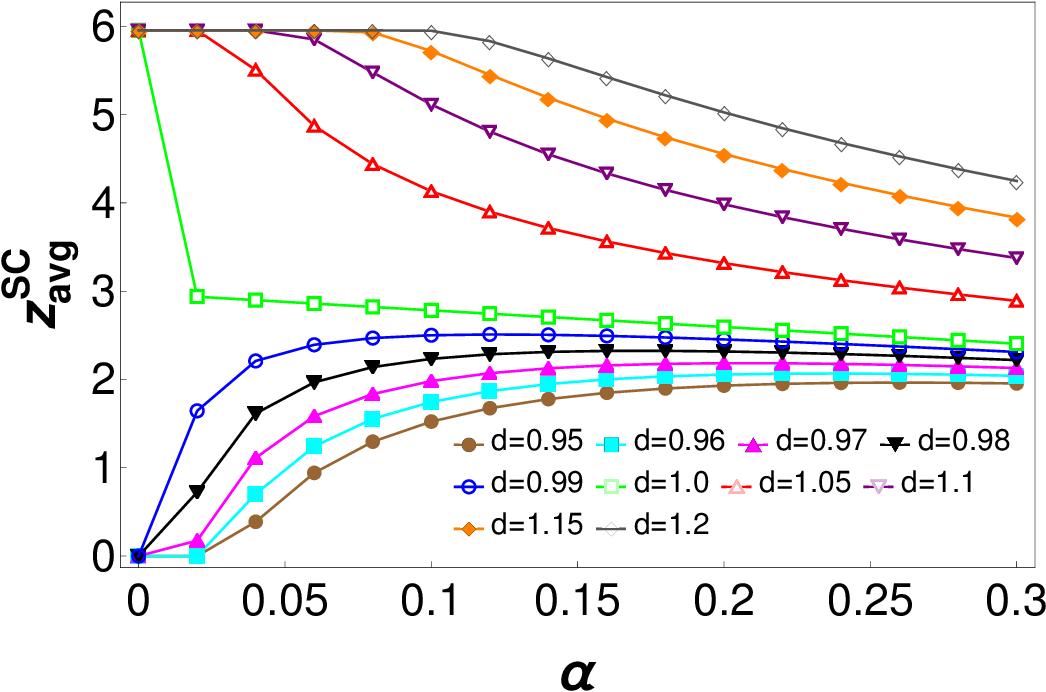}}
     \subfigure[]{\includegraphics[scale=0.5]{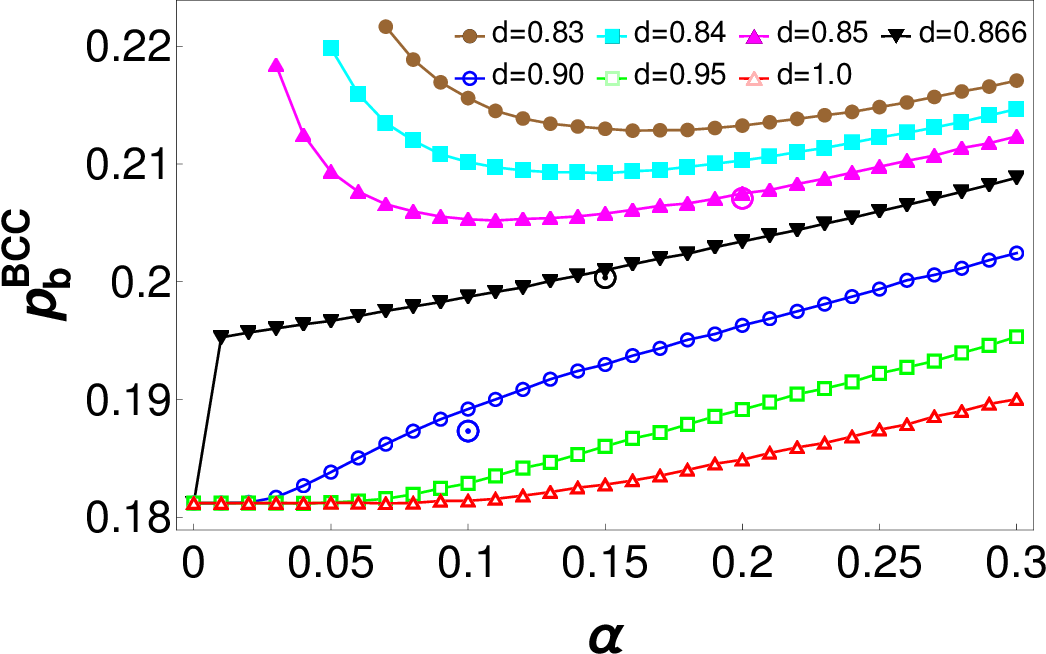}}\hspace{0.2cm}
     \subfigure[]{\includegraphics[scale=0.47]{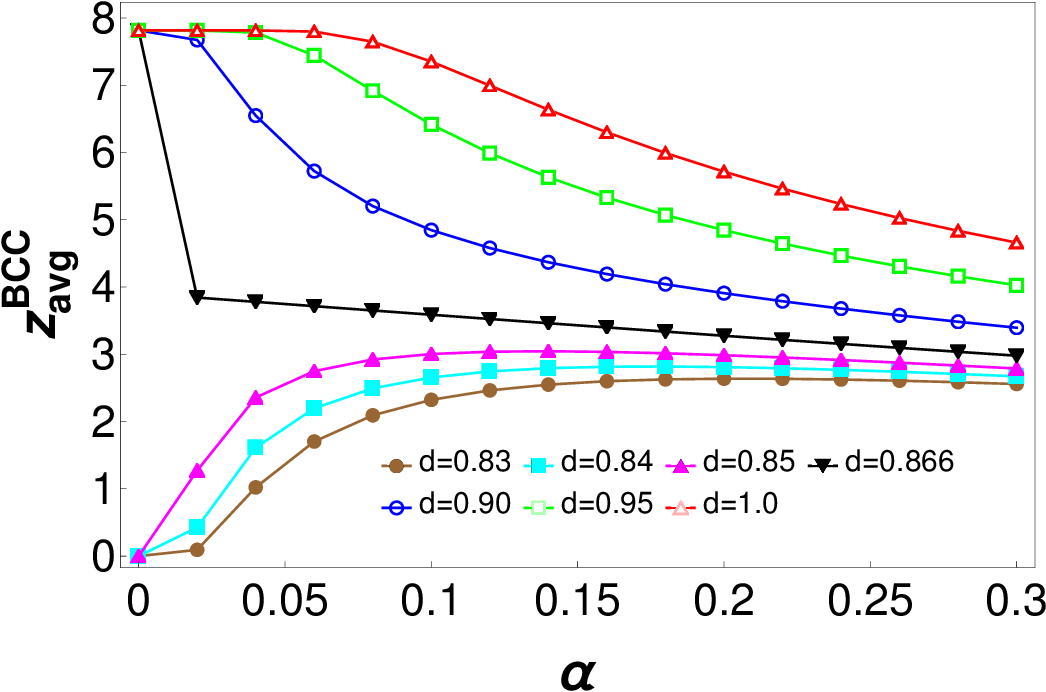}}
\caption{(a) Variation of bond percolation threshold $(p_\mathrm{b})$ with distortion parameter $(\alpha)$ for distorted SC lattice of linear size $L=2^{7}$. Ten different curves correspond to different values of $d$. (b) Corresponding variation of the average coordination number $z_\mathrm{avg}(\alpha)$ satisfying $\delta\le d$ for DSC lattice. Curves for same values of $d$ are represented by same colors in (a) and (b). (c) Variation of bond percolation threshold $(p_\mathrm{b})$ with distortion parameter $(\alpha)$ for distorted BCC lattice of linear size $L=2^{6}$. Seven different curves correspond to different values of $d$. (d) Corresponding variation of the average coordination number $z_\mathrm{avg}(\alpha)$ satisfying $\delta\le d$ for DBCC lattice. Curves for same values of $d$ are represented by same colors in (c) and (d). Each data point of panels (a) and (c) represents an average over $1000$ independent realizations and of panels (b) and (d) represents an average over $100$ independent realizations. The data points are joined by lines as a guide to the eye. The corresponding percolation thresholds in the thermodynamic limit for some selected values of $d$ and $\alpha$ are shown by the symbol $\odot$ in the same colors.}
\label{fig:pb_alpha}
\end{figure*}

\section{Variation of Bond Percolation Threshold}\label{VBPT}
This section examines how the bond percolation threshold for distorted simple cubic (DSC) and distorted body-centered cubic (DBCC) lattices depends on the distortion parameter $\alpha$ and the connection threshold $d$. For reference, the established bond percolation thresholds of the corresponding infinite, undistorted lattices are approximately $p_\mathrm{b}=0.2488$ for the simple cubic (SC) lattice and $p_\mathrm{b}=0.18025$ for the body-centered cubic (BCC) lattice \cite{Lorenz}.

The variation of the bond percolation threshold $p_\mathrm{b}$ with $\alpha$ for different values of $d$ is shown in Fig.~\ref{fig:pb_alpha}.(a) and Fig.~\ref{fig:pb_alpha}.(c). Each data point was obtained from simulations on lattices of size $L=2^{7}$ for the DSC lattice and $L=2^{6}$ for the DBCC lattice, and represents an average over $1000$ independent realizations. The behavior of the bond percolation threshold differs markedly depending on whether the connection threshold $d$ is greater than or less than the lattice constant.

For \(d\ge 1\), the percolation threshold$(p_\mathrm{b})$ increases monotonically with the distortion parameter $\alpha$ for both lattice types. As the distortion increases, the Euclidean distance $\delta$ between neighboring sites grows for some bonds. When this distance exceeds the connection threshold $d$, those bonds are effectively removed. The resulting reduction in connectivity makes the formation of a spanning cluster more difficult, thereby increasing the percolation threshold. In the limit of very small $\alpha$ with \(d\ge 1\), the values of $p_\mathrm{b}$ approach those of the undistorted SC and BCC lattices.

For $d<1$, the dependence of the percolation threshold on $\alpha$ becomes non-monotonic for both lattices. At very small distortion ($\alpha \approx 0$), the bond lengths exceed the connection threshold, so no bonds are present and percolation does not occur. As $\alpha$ increases, random displacements bring some neighboring sites closer together, allowing certain bond distances to fall below $d$. This is why all the curves corresponding to $d<1$ originate at a finite, nonzero value of $\alpha$. This enhanced local connectivity initially lowers the percolation threshold. However, at larger values of $\alpha$, the average distance between neighboring sites increases again, reducing the number of available bonds and causing the percolation threshold to rise.

When $d=1$, a sharp increase in $p_\mathrm{b}$ is observed for both lattices even for a small increase in $\alpha$. In this case, the percolation threshold jumps abruptly from the value corresponding to the undistorted lattice to a significantly higher value.

To better understand the behavior of $p_\mathrm{b}(\alpha)$, we analyze the average coordination number $z_\mathrm{avg}(\alpha)$ for DSC and DBCC lattices of sizes $L=2^{7}$ and $L=2^{6}$, respectively. For $d>1$ (see Fig.~\ref{fig:pb_alpha}.(b) and Fig.~\ref{fig:pb_alpha}.(d)), the average coordination number decreases with increasing $\alpha$. At $\alpha=0$, all curves yield $z_\mathrm{avg}=6$ for the DSC lattice and $z_\mathrm{avg}=8$ for the DBCC lattice, consistent with their undistorted structures. 

For $d<1$, the average coordination number starts from zero at $\alpha=0$ and increases with increasing distortion. In contrast, for $d=1$, $z_\mathrm{avg}$ exhibits a discontinuous drop from $6$ to $3$ for the DSC lattice and from $8$ to $4$ for the DBCC lattice. The observed discontinuous jump can be attributed to variations in bond length. Since bonds are equally likely to be stretched or compressed, approximately half of the neighboring bonds associated with a given bond fail to satisfy the bond occupation criterion. Consequently, the effective average coordination number is reduced by roughly one half. A decrease in the average coordination number with increasing $\alpha$ indicates a reduction in the number of nearest neighbors. Consequently, a larger fraction of bonds must be occupied to achieve a spanning cluster, leading to an increase in the bond percolation threshold $p_\mathrm{b}$

\begin{figure*}
	  \subfigure[]{\includegraphics[scale=0.47]{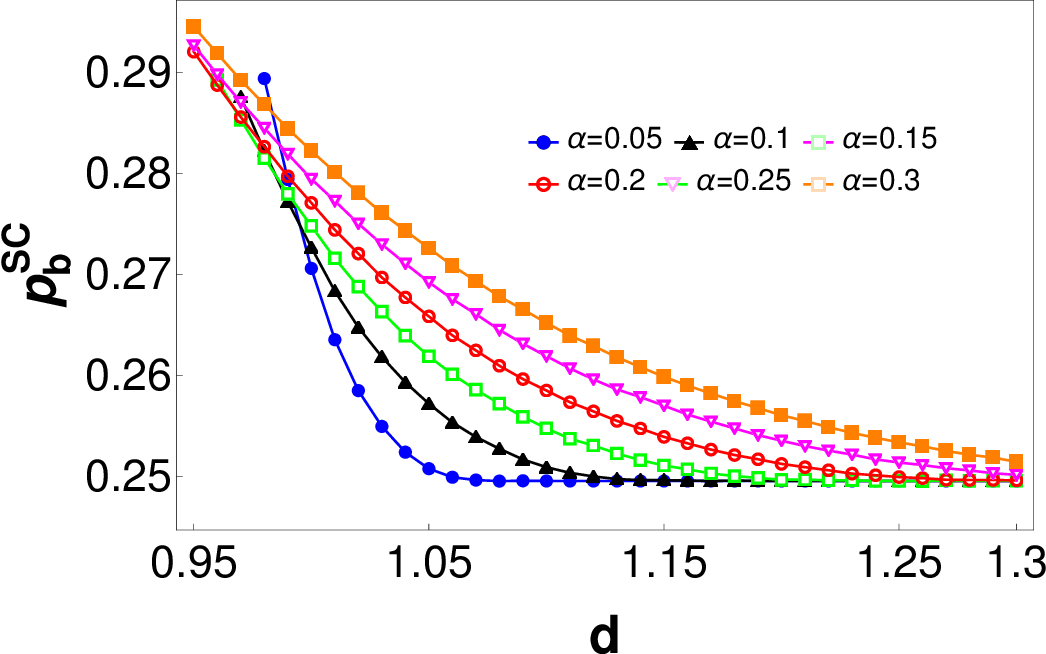}}\hspace{0.2cm}
      \subfigure[]{\includegraphics[scale=0.47]{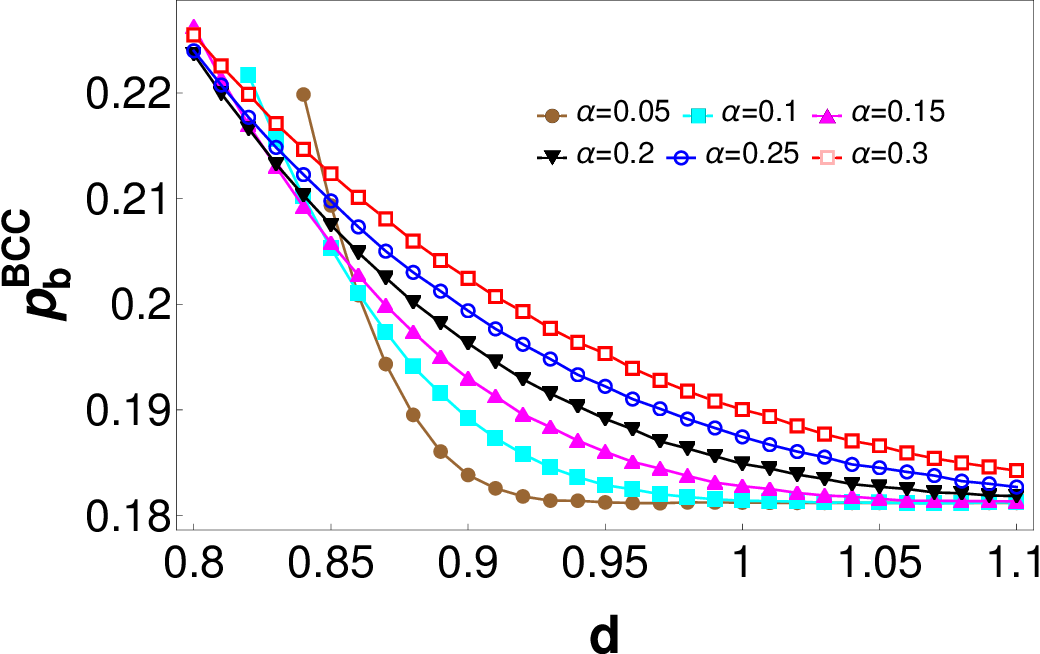}}
\caption{(a) Variation of bond percolation threshold $(p_\mathrm{b})$ with connection threshold $(d)$ for distorted SC lattice of linear size $L=2^{7}$. Six different curves correspond to different values of $\alpha$. (b) Variation of bond percolation threshold $(p_\mathrm{b})$ with connection threshold $(d)$ for distorted BCC lattice of linear size $L=2^{6}$. Six different curves correspond to different values of $\alpha$. Each data point represents an average over $1000$ independent realizations. The data points are joined by lines as a guide to the eye.}
\label{fig:pb_d}
\end{figure*}

Fig.~\ref{fig:pb_d} shows the variation of the bond percolation threshold ($p_\mathrm{b}$), as a function of the connection threshold ($d$), for different values of the distortion parameter $\alpha$. Each data point represents an average over $1000$ independent realizations, using lattice sizes $L=2^{7}$ for the DSC lattice and $L=2^{6}$ for the DBCC lattice. The percolation threshold $p_\mathrm{b}$ decreases monotonically with increasing $d$. This behavior arises because a larger connection threshold increases the probability that the Euclidean bond length$(\delta)$ is smaller than $d$, leading to a higher fraction of occupied bonds. As a result, the formation of a spanning cluster becomes more probable. In the limit of large $d$, the percolation threshold $p_\mathrm{b}$ approaches the corresponding undistorted values of the SC and BCC lattices. Additionally, the curves exhibit a crossing point in the regime $d<1$, indicating a nontrivial interplay between lattice distortion and connectivity.

\begin{table}
\centering
\begin{tabular}{|c|c|c|c|c|}
\hline 
\makecell{Lattice type} & $d$ & $\alpha$
& \makecell{$p_\mathrm{b}^{\infty}(\alpha,d)$}  \\ 
\hline 
\makecell{Distorted\\Simple cubic} & \makecell{0.97\\1.0\\1.1} & \makecell{0.1\\0.05\\0.2} & \makecell{0.2865(2)\\0.2702(1)\\0.2577(2)}\\ 
\hline 
\makecell{Distorted\\Body-centered cubic} & \makecell{0.85\\0.90\\$\frac{\sqrt{3}}{2}$} & \makecell{0.2\\0.1\\0.15} & \makecell{0.2070(8)\\0.1873(8)\\0.2003(1)} \\ 
\hline 
\end{tabular} 
\caption{Precise values of $p_\mathrm{b}(\alpha,d)$ obtained from Binder cumulant.}
\label{Tab:precise}
\end{table}

\begin{figure*}
\subfigure[]{\includegraphics[scale=0.4]{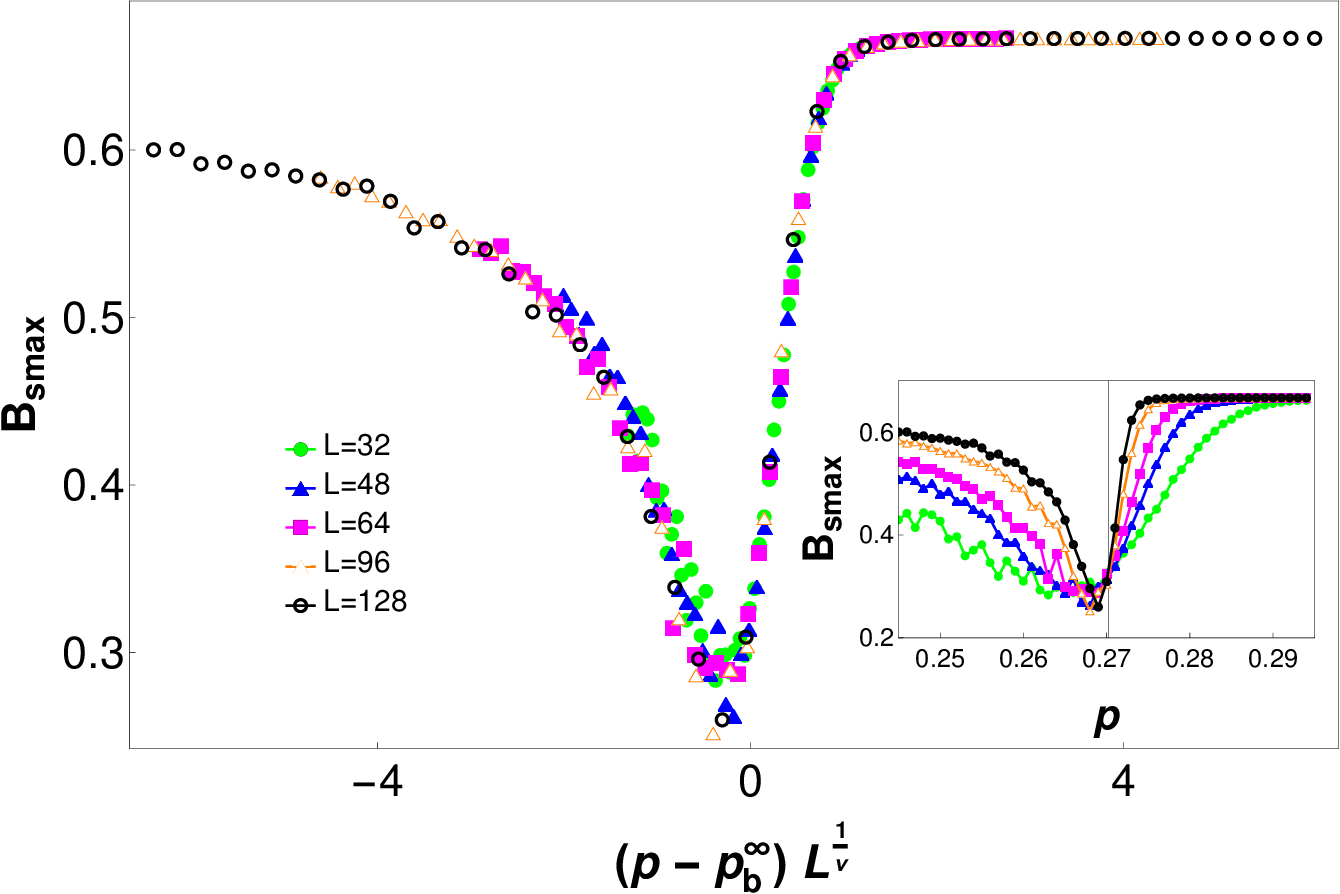}}
\hspace{0.1cm}
\subfigure[]{\includegraphics[scale=0.4]{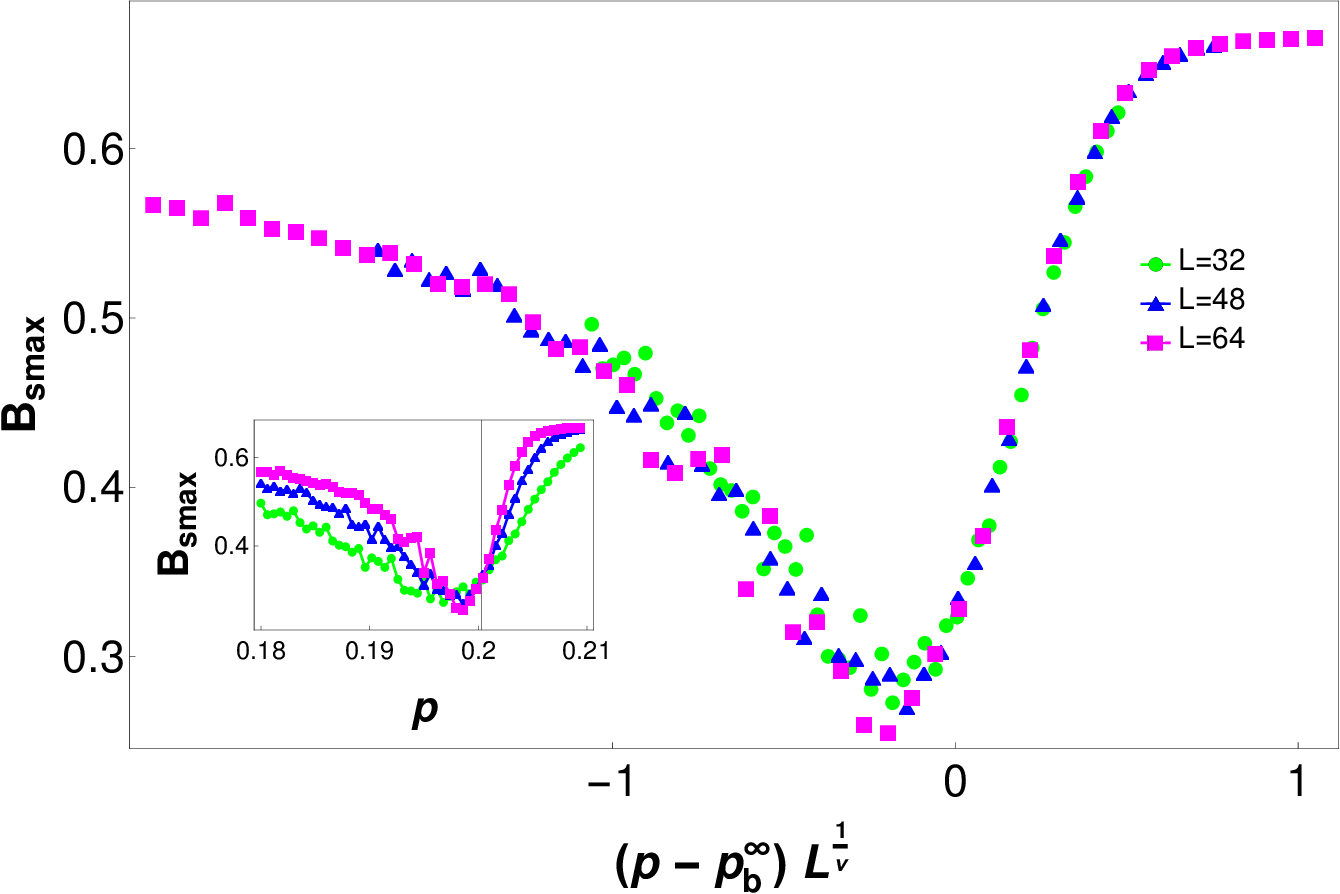}}
\caption{Determination of the bond percolation threshold in the thermodynamic limit (insets) and demonstration of universality through finite-size scaling (main panels). The threshold $p_\mathrm{b}^{\infty}$ is obtained from the mean of the intersection points of Binder cumulant curves for different lattice sizes, as indicated by the vertical lines in the insets, and the corresponding values are listed in Table~\ref{Tab:precise}. The horizontal axis is then rescaled as $(p - p_\mathrm{b}^{\infty})L^{1/\nu}$, using the standard three-dimensional percolation exponent $\nu=0.8762$, resulting in a good data collapse of the Binder cumulant. Each data point represents an average over $10^{4}$ independent realizations, and lines are drawn as a guide to the eye. Panel (a) shows results for DSC lattices of sizes $L = 32, 48, 64, 96,$ and $128$ with $d = 1.0$ and $\alpha = 0.05$, while panel (b) presents corresponding results for DBCC lattices of sizes $L = 32, 48,$ and $64$ with $d = \sqrt{3}/2$ and $\alpha = 0.15$.}
\label{fig:sc_binder}
\end{figure*}

\section{Thresholds in the thermodynamic limit and universality}
In this section, we determine precise estimates of the bond percolation threshold of DSC and DBCC lattices in the thermodynamic limit for selected combinations of $\alpha$ and $d$. These estimates are obtained from the intersection points of Binder cumulant curves corresponding to different lattice sizes at fixed values of $\alpha$ and $d$. The mean of these intersection points provides an estimate of the percolation threshold in the thermodynamic limit.

The Binder cumulant is defined as
\begin{equation}
B_\mathrm{smax}(p,L)=1-\frac{\langle [S_\mathrm{max}(p,L)]^4\rangle}{3\langle [S_\mathrm{max}(p,L)]^2\rangle^2},
\end{equation}
where $S_\mathrm{max}(p,L)$ denotes the size of the largest cluster in a lattice of linear size $L$ at bond occupation probability $p$, and $\langle \cdot \rangle$ represents an average over independent realizations. In our simulations, all quantities are averaged over $10^{4}$ independent configurations.

The intersections of the Binder cumulant curves for different lattice sizes are shown in the insets of Fig.~\ref{fig:sc_binder}(a) for the DSC lattice and Fig.~\ref{fig:sc_binder}(b) for the DBCC lattice. The resulting estimates of the bond percolation threshold $p_\mathrm{b}^{\infty}(\alpha,d)$ are summarized in Table~\ref{Tab:precise}. These values are also indicated by the symbol $\odot$ in Figs.~\ref{fig:pb_alpha}(a) and \ref{fig:pb_alpha}(c). As expected, the thermodynamic thresholds are close to those obtained for finite lattice sizes, with only weak finite-size dependence.

To further validate the accuracy of the obtained thresholds, we perform a finite-size scaling analysis, as shown in the main panels of Fig.~\ref{fig:sc_binder}. The horizontal axis is rescaled as $(p - p_\mathrm{b}^{\infty})L^{1/\nu}$ using the estimated values of $p_\mathrm{b}^{\infty}$ and the standard three-dimensional percolation exponent $\nu = 0.8762$ \cite{Brzeski_2022}. This rescaling leads to a satisfactory collapse of the Binder cumulant data for different system sizes onto a single curve.

This data collapse provides strong support for the accuracy of the estimated thermodynamic thresholds and indicates that the scaling behavior is consistent with that of standard three-dimensional percolation. Although we do not independently determine all critical exponents, the observed scaling behavior and the absence of anomalies suggest that the universality class remains unchanged in the presence of distortion. This conclusion is in agreement with our earlier study of site percolation in DSC lattices, where the transition was also found to belong to the standard three-dimensional percolation universality class \cite{Sayantan2}.

\section{Critical Connection Threshold} \label{CCT}
Our earlier results demonstrate that, for a given lattice distortion, the formation of a spanning cluster occurs only when the connection threshold $d$ reaches a specific value. This observation indicates that, at a fixed distortion parameter $\alpha$, a minimum value of connection threshold is necessary to achieve global connectivity. We define this minimum value of connection threshold as the critical connection threshold $d_\mathrm{c}$, which depends explicitly on $\alpha$. The numerical procedure used to estimate $d_\mathrm{c}$ is described as follows :

\begin{figure*}
	  \subfigure[]{\includegraphics[scale=0.55]{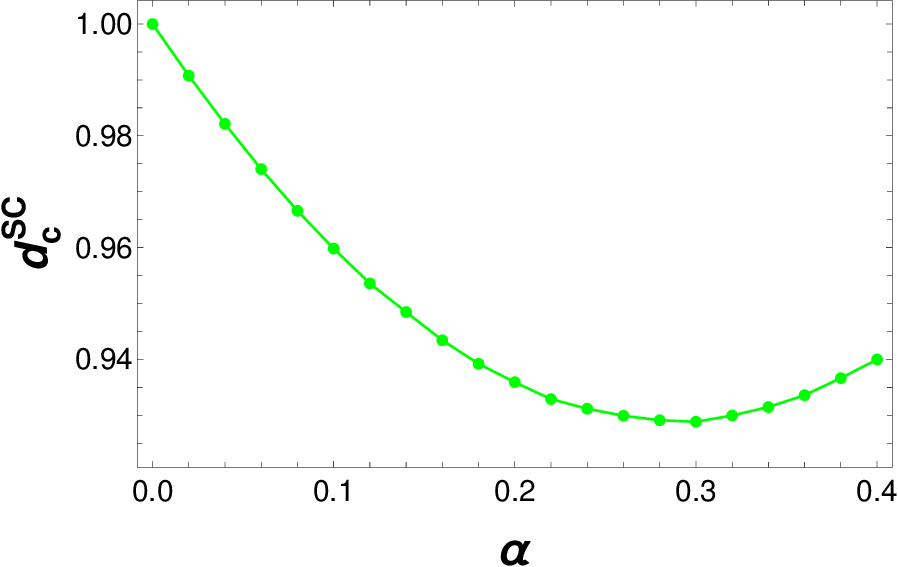}}\hspace{0.3cm}
     \subfigure[]{\includegraphics[scale=0.55]{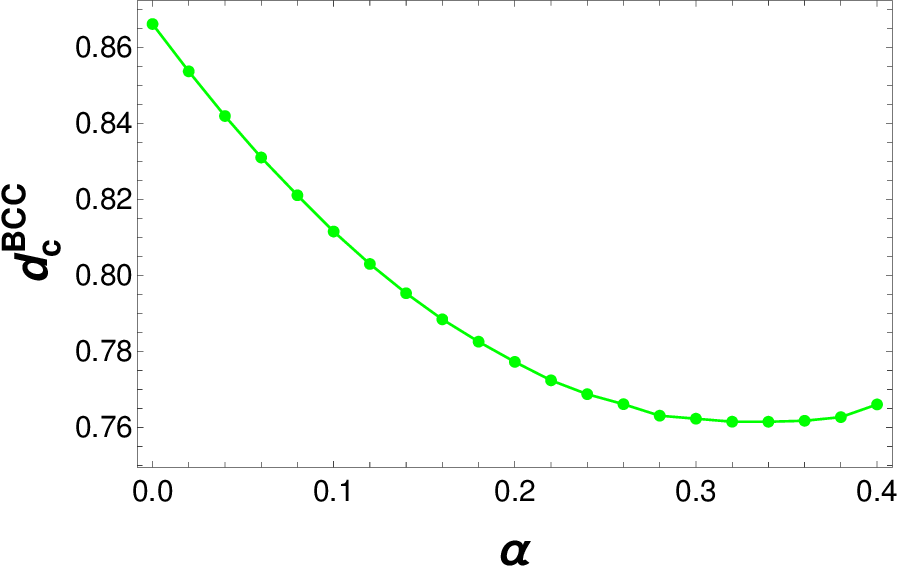}}
     
\caption{(a) Variation of critical connection threshold $(d_\mathrm{c})$ with distortion parameter $(\alpha)$ for distorted SC lattice of linear size $L=2^{7}$. (b) Variation of critical connection threshold $(d_\mathrm{c})$ with distortion parameter $(\alpha)$ for distorted BCC lattice of linear size $L=2^{6}$. Each data point represents an average over $1000$ independent realizations. The data points are joined by lines as a guide to the eye.}
\label{fig:dc}
\end{figure*}

\begin{enumerate}
    \item To begin with, a distorted lattice is formed with a fixed value of \(\alpha\) and \(d\).
    \item All the bonds satisfying $\delta \le d$ are occupied.
    \item Initially, the connection threshold $d$ is kept at a low value. The value is so low that spanning is not possible even though the condition (ii) is applied.
    \item $d$ is now slowly increased. Spanning is checked for each value of $d$.
    \item The first value of d, when spanning is obtained, is noted.
    \item Hence a new configuration is generated for the same value of $\alpha$ and whole process is repeated.
    \item An average of $10$ such values of $d$ is identified as the critical connection threshold $d_\mathrm{c}$
\end{enumerate}

Fig.~\ref{fig:dc}.(a) and Fig.~\ref{fig:dc}.(b) illustrates the dependence of the critical connection threshold $d_\mathrm{c}$ on the distortion parameter $\alpha$ for distorted simple cubic (DSC) and distorted body-centered cubic (DBCC) lattices respectively. For the DSC lattice (see Fig.~\ref{fig:dc}.(a)), the undistorted case $(\alpha=0)$ yields $d_\mathrm{c}=1$ , equal to the lattice constant. As $\alpha$ increases, $d_\mathrm{c}$ decreases and attains a minimum value of $0.9288$ at $\alpha=0.3$, beyond which it increases monotonically. A similar behavior is observed for the DBCC lattice (see Fig.~\ref{fig:dc}.(b)). At $\alpha=0$, the critical connection threshold equals the nearest-neighbor distance of the BCC lattice, $d_\mathrm{c}=\frac{\sqrt{3}}{2}$. With increasing distortion, $d_\mathrm{c}$ initially decreases, reaching a minimum value of approximately $0.7614$ at $\alpha=0.34$, and subsequently increases for larger $\alpha$.

Together with our earlier results for square and triangular lattices, these findings indicate that, except for the square lattice, $d_\mathrm{c}$ generically exhibits a nonmonotonic dependence on lattice distortion. This behavior can be attributed to the coordination number exceeding four in these lattices. In contrast, for the square lattice, which has exactly four nearest neighbors, $d_\mathrm{c}$ increases monotonically with $\alpha$

\begin{figure*}
      \subfigure[]{\includegraphics[scale=0.55]{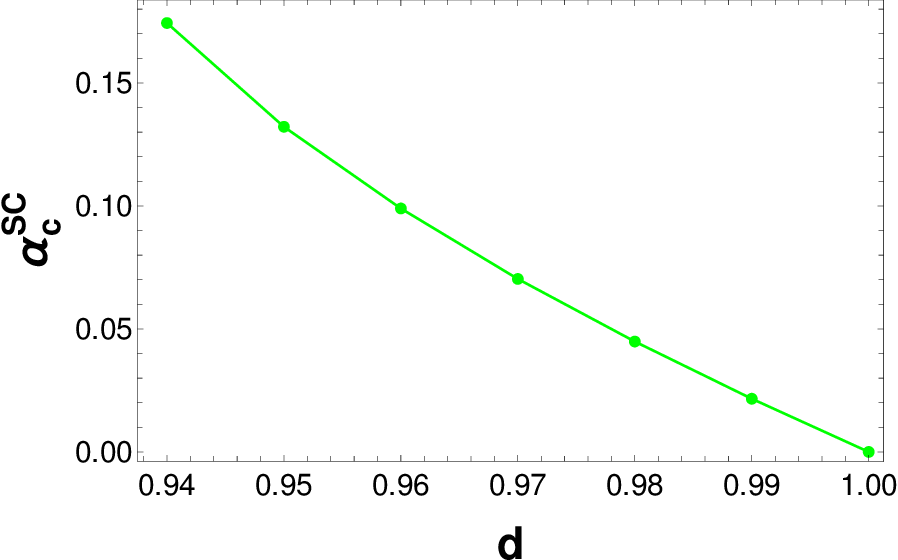}}
     \subfigure[]{\includegraphics[scale=0.55]{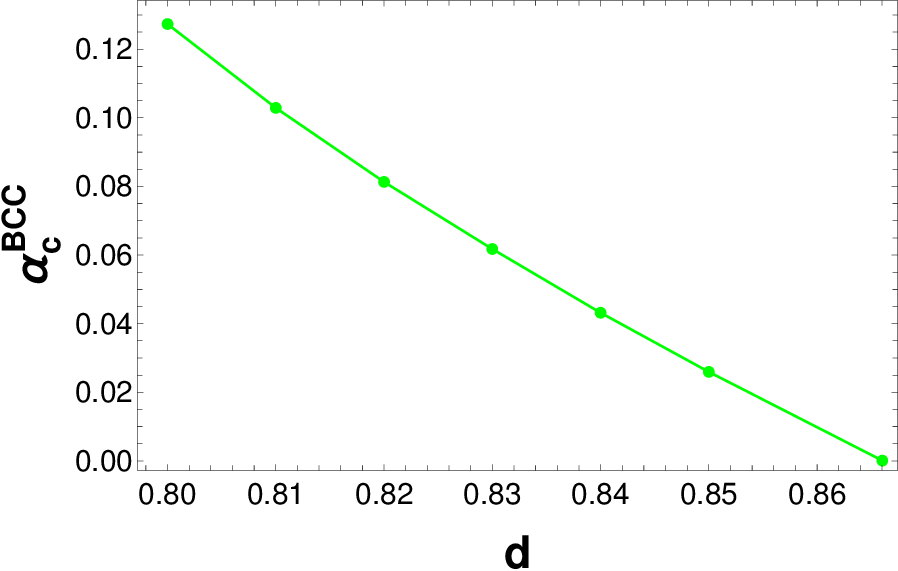}}
     
\caption{(a) Variation of critical distortion parameter ($\alpha_\mathrm{c}$) with connection threshold $(d)$ for distorted SC lattice of linear size $L=2^{7}$. (b) Variation of critical distortion parameter ($\alpha_\mathrm{c}$) with connection threshold $(d)$ for distorted BCC lattice of linear size $L=2^{6}$. Each data point represents an average over $1000$ independent realizations. The data points are joined by lines as a guide to the eye.}
\label{fig:alphac}
\end{figure*}

\section{Critical Distortion Parameter}\label{CDP}
For $d<1$, spanning is not achieved (see Fig.~\ref{fig:pb_alpha}.(a) and Fig.~\ref{fig:pb_alpha}.(c)) in the absence of distortion $(\alpha=0)$. In this regime, all curves originate at a finite value of $\alpha$, indicating that a minimum amount of lattice distortion is required to form a spanning cluster. This minimum distortion defines the critical distortion parameter $\alpha_\mathrm{c}$. Fig.~\ref{fig:alphac}.(a) and Fig.~\ref{fig:alphac}.(b) show the dependence of the critical distortion parameter $\alpha_\mathrm{c}$ on the connection threshold $d$ for DSC and DBCC lattices, respectively. Fig.~\ref{fig:alphac}.(a) and Fig.~\ref{fig:alphac}.(b) further demonstrate that $\alpha_\mathrm{c}$ decreases monotonically with increasing $d$, implying that larger connection thresholds reduce the minimum distortion required to achieve global connectivity. The numerical estimation of \(\alpha_\mathrm{c}\) proceeds in the following manner:

\begin{enumerate}
    \item To begin with, a distorted lattice is formed with a fixed value of \(\alpha\) and \(d\).
    \item All the bonds satisfying $\delta \le d$ are occupied.
    \item Initially, the distortion parameter \(\alpha\) is kept at a low value. The value is so low that spanning is not possible even though the condition (ii) is applied.
    \item \(\alpha\) is now slowly increased. Spanning is checked for each value of $\alpha$.
    \item The first value of \(\alpha\), when spanning is obtained, is noted.
    \item Hence a new configuration is generated for the same value of $d$ and whole process is repeated.
    \item An average of $10$ such values of $\alpha$ is identified as the critical distortion parameter $\alpha_\mathrm{c}$.
\end{enumerate}
\section{Discussion: Novelty, Physical Insights, and Comparison with Earlier Works}

In this section, we clarify the novelty of the present work, its relation to our earlier studies on site and bond percolation in distorted lattices, and the physical insights that emerge from the present analysis.

\subsection{Relation to earlier works}

The present study builds upon our earlier investigations of percolation in distorted two- and three-dimensional lattices \cite{Sayantan1, Sayantan2, Sayantan3, Bishnu1}, where the effects of geometric distortion and a distance-dependent connectivity criterion were studied. In particular, the impact of distortion on site percolation in distorted simple cubic (DSC) lattices has been studied in detail, along with a comprehensive analysis of critical behavior \cite{Sayantan2}.

In contrast, the present work addresses bond percolation in three-dimensional distorted lattices. Although the underlying distortion mechanism is identical, the process of cluster formation is fundamentally different. In the case of site percolation, all sites are initially empty and all bonds that satisfy the connection criterion are initially occupied. The sites are then randomly occupied until the threshold is reached. In contrast, bond percolation considers a different initial configuration with all occupied sites and all empty bonds. In this case, only the bonds satisfying the connection criterion are eligible for occupation. The purpose of the present study is to assess the impact of this additional restriction on the bond percolation threshold of distorted three-dimensional lattices.

Furthermore, the inclusion of the distorted body-centered cubic (DBCC) lattice extends the scope of the analysis beyond the distorted simple cubic (DSC) geometry considered previously. This allows us to systematically investigate the influence of lattice structure and coordination number on percolation in the presence of distortion.

In summary, earlier studies established the role of geometric distortion and distance-dependent connectivity in determining percolation behavior, particularly for site percolation in DSC lattices. In contrast, the present work provides new results for bond percolation in distorted lattices, extends the analysis to the DBCC geometry, and reveals additional features such as the dependence of crossover behavior on the nearest-neighbor distance of the underlying lattice.

\subsection{Novel contributions of the present work}

The main novel contributions of this work can be summarized as follows:

(i) We present a systematic study of bond percolation on distorted three-dimensional lattices, which has not been addressed in any of our earlier works or in works of others, to our knowledge.

(ii) We obtain the bond percolation threshold $p_\mathrm{b}(\alpha,d)$ of finite DSC and DBCC lattices over a wide range of connection threshold and distortion parameter. A direct comparison with earlier site-percolation results shows that bond percolation is systematically less sensitive to geometric distortion than site percolation in DSC lattices. While both models exhibit qualitatively similar trends with respect to the distortion parameters, the variation in the bond percolation threshold is noticeably weaker. This difference arises from the distinct mechanisms of cluster formation in the two cases, as discussed above.

(iii) For both DSC and DBCC lattices, the nature of variation of the bond percolation threshold changes altogether when the connection threshold becomes less than the nearest-neighbor distance of the corresponding regular lattice. In particular, this crossover occurs at $d=1$ for the DSC lattice and at $d=\sqrt{3}/2$ for the DBCC lattice, corresponding to their respective nearest-neighbor distances. A discontinuous jump is seen for both lattices when the connection threshold and the nearest-neighbor distance are equal.

(iv) We identify the corresponding critical values of the connection threshold $d_\mathrm{c}$ and the distortion parameter $\alpha_\mathrm{c}$, below which no spanning configuration can be obtained even when all bonds satisfying the connection criterion are occupied. Notably, these quantities encode genuinely new physical information and do not merely provide alternative representations of the trends in $p_\mathrm{b}(\alpha, d)$. The determination of the curves $d_\mathrm{c}(\alpha)$ and $\alpha_\mathrm{c}(d)$ required independent numerical implementation.

(v) By comparing DSC and DBCC lattices, we demonstrate how lattice geometry and coordination number influence the bond percolation properties under distortion.

(vi) We perform a finite-size scaling analysis using Binder cumulants and demonstrate data collapse consistent with the standard three-dimensional percolation exponent $\nu$, thereby supporting the numerical accuracy and consistency with the expected universality class.

\subsection{Physical insights}

The results of this study provide several physical insights into percolation in geometrically disordered systems.

First, an important observation emerging from the present analysis is the identification of a natural geometric scale that governs the behavior of the system. In the case of the DSC lattice, a qualitatively distinct set of curves is observed when $d<1$, where unity corresponds to the nearest-neighbor distance of the undistorted simple cubic lattice. However, for the DBCC lattice, a similar qualitative change occurs when $d<\sqrt{3}/2$, which is precisely the nearest-neighbor distance of the undistorted body-centered cubic lattice. This clearly indicates that the relevant control parameter is its value relative to the nearest-neighbor distance of the corresponding undistorted lattice. When $d$ falls below this characteristic length scale, the effective connectivity is significantly altered, leading to a distinct percolation behavior. This observation provides a unified geometric interpretation of the results across different lattice types.

Second, the dependence of $p_\mathrm{b}$ on the parameters $\alpha$ and $d$ reflects the competition and interplay between these two parameters. When the connection threshold is greater than the undistorted nearest-neighbor distance, the effective connectivity of the lattice is reduced as distortion increases, leading to higher percolation thresholds. However, distortion can sometimes enhance connectivity and reduce $p_\mathrm{b}$, as seen in the low-$\alpha$ regime of the curves for which $d$ is less than the nearest-neighbor distance of the regular lattice (see Fig.~\ref{fig:pb_alpha}).

Third, the comparison between DSC and DBCC lattices highlights the role of coordination number. The higher coordination number of the BCC lattice leads to enhanced connectivity, making the DBCC lattice more resilient to distortion compared to the DSC lattice as reflected in lower threshold values.

Fourth, the observation of a consistent data collapse using the standard three-dimensional exponent $\nu$ indicates that geometric distortion does not alter the universality class of the transition. Instead, distortion primarily shifts the location of the critical point without affecting the underlying critical behavior.

\subsection{Summary of significance}

In summary, the present work extends the framework of geometrically distorted lattices to bond percolation and provides a systematic comparison between different lattice geometries. The results demonstrate how distortion and lattice structure jointly influence percolation thresholds while preserving universal critical behavior. While qualitatively similar trends in $p_\mathrm{b}(\alpha,d)$ are expected to persist for other lattice structures with distance-based connectivity, the specific quantitative behavior depends on several factors, including (i) the lattice geometry, (ii) the coordination number and its variation with distortion, and (iii) the type of percolation (bond or site). These findings contribute to a broader understanding of percolation in disordered systems and complement earlier studies on site percolation in similar settings.

\section{Conclusions}

We study the impact of geometric distortion on the bond percolation threshold in simple cubic and body-centered cubic lattices of lattice constant $1$. The distortion is introduced while preserving the underlying lattice topology through a tunable parameter $\alpha$ in the following manner. The lattice sites are dislocated within a cube of length $\alpha$ around their regular lattice positions. As a result, the bond length $\delta$ is no longer constant. A bond can only be occupied if its length $\delta$ is less than or equal to a prescribed threshold $d$, called the connection threshold. Through large-scale Monte Carlo simulation and finite-size scaling analysis, we examine how the bond percolation threshold of finite and infinite lattices varies with distortion. We summarize our findings below.

\begin{enumerate}

\item As shown in Fig.~\ref{fig:pb_alpha}(a) and (c), when the connection threshold is greater than the nearest-neighbor distance of the undistorted lattice ($d>1$ for SC and $d>\sqrt{3}/2$ for BCC), the bond percolation threshold $p_\mathrm{b}$ starts from its established value for the corresponding regular lattice and then increases monotonically with the distortion parameter $\alpha$. This implies that spanning becomes progressively more difficult as distortion increases.

\item The monotonicity breaks when the connection threshold is less than the regular nearest-neighbor distance: $p_\mathrm{b}$ initially decreases with $\alpha$, reaches a minimum, and then increases steadily, as evident from Fig.~\ref{fig:pb_alpha}(a) and (c).

\item When the connection threshold is equal to the regular nearest-neighbor distance ($d=1$ for SC and $d=\sqrt{3}/2$ for BCC), a sharp rise in $p_\mathrm{b}$ is observed as $\alpha$ changes from $0$ to any infinitesimal value, followed by a steady approximately linear increase.

\item All these variations of $p_\mathrm{b}$ can be explained from the plots (in Fig.~\ref{fig:pb_alpha}(b) and (d))of the average coordination number $z_\mathrm{avg}(\alpha)$, which exhibit corresponding inverse trends.

\item Bond percolation thresholds in the thermodynamic limit, evaluated through the intersections of the Binder cumulant, demonstrate that the finite-size effects are minimal and confirm that the variations of $p_\mathrm{b}$ for finite-sized lattices correctly reflect the dependence of the bond percolation threshold on distortion. The values are listed in Table~\ref{Tab:precise} and highlighted in Fig.~\ref{fig:pb_alpha}(a) and (c) with distinct symbol.

\item To gain further insight into the mechanism of connectivity in the distorted three-dimensional lattices, we determine the critical value of the connection threshold $d_\mathrm{c}$ required to achieve spanning when all the allowed bonds are occupied. We find a smooth non-monotonic dependence for both lattices : an initial decline, formation of a minimum, followed by an increment, as shown in Fig.~\ref{fig:dc}.

\item Similarly, for connection threshold less than the regular nearest-neighbor distance, we evaluate a critical value of the distortion parameter $\alpha_\mathrm{c}$ required for spanning when all the allowed bonds are occupied. Fig.~\ref{fig:alphac} shows that $\alpha_\mathrm{c}$ decreases steadily for both lattices.
\end{enumerate}
\section*{Acknowledgments}
This work is funded by Anusandhan National Research Foundation (ANRF), Department of Science and Technology, Government of India. The project file number is SUR/2022/002345. Sayantan Mitra gratefully acknowledges financial support through a National Post Doctoral Fellowship from  ANRF under  project file no. PDF/2023/002952. The authors thank Abdur Rashid Miah for useful discussion. The computation facilities availed at the Department of Physics, University of Gour Banga, Malda are gratefully acknowledged.
\section*{References}
\providecommand{\newblock}{}

\end{document}